\documentclass[printer]{aa}

\usepackage{graphicx}
\usepackage{txfonts}
\usepackage{bm}
\usepackage{xcolor}

\usepackage{longtable} 
\usepackage{pdflscape} 
\usepackage{multicol}
\usepackage{multirow}
\usepackage{booktabs}

\begin{document}

   \title{Nitrogen- and nitrogen-oxygen-bearing organic molecules in comet 67P/Churyumov-Gerasimenko: An untargeted investigation}
   \titlerunning{N- and NO-bearing organic molecules in comet 67P}

\author{N. H{\"a}nni\inst{1} \and K. Altwegg\inst{1} \and D. Baklouti\inst{2} \and M. Combi\inst{3} \and S. A. Fuselier\inst{4,5} \and J. De Keyser\inst{6} \and D. R. Müller\inst{1} \and M. Rubin\inst{1} \and S. F. Wampfler\inst{7}}

\institute{Space Research \& Planetary Sciences, Physics Institute, University of Bern, Sidlerstrasse 5, CH-3012 Bern, Switzerland.\\ \email{nora.haenni@unibe.ch}
\and
Institut d'Astrophysique Spatiale, Université Paris-Saclay, CNRS, Orsay, France.
\and
Department of Climate and Space Sciences and Engineering, University of Michigan, Ann Arbor, MI, USA.
\and
Space Science Division, Southwest Research Institute, San Antonio, TX, USA.
\and
Department of Physics and Astronomy, The University of Texas at San Antonio, San Antonio, TX, USA.
\and
Royal Belgian Institute for Space Aeronomy, BIRA-IASB, Brussels, Belgium.
\and
Center for Space and Habitability, University of Bern, Gesellschaftsstrasse 6, CH-3012 Bern, Switzerland.}

\date{Received XXX / Accepted YYY}

\abstract
{Comets provide a unique window into the history of the Solar System  as they carry some of the best-preserved material and make it available to in situ exploration. A milestone in comet studies was the European Space Agency's Rosetta mission, which, for the first time, rendezvoused with a comet, namely 67P/Churyumov-Gerasimenko (67P), and studied it from a close range for two years. Amongst other unexpected insights, data from this mission show that comets contain a surprisingly large portion of organics, both in the refractory and the icy phases. For this work, we  evaluated high-resolution mass spectra collected in comet 67P's inner coma by Rosetta's ROSINA-Double Focusing Mass Spectrometer (DFMS). In unprecedented detail, we  investigated the N- and NO-bearing cometary complex organic molecules (COMs) of the general sum formula C$_n$H$_m$N and C$_n$H$_m$NO, where $n$ and $m$ are the stoichiometric coefficients of carbon and hydrogen. Our discussion-driven approach combines the empirical concept of Occam's razor with knowledge from studies of relevant astrophysical environments and constraints expected from naive bottom-up assembly of molecules. We present an exemplary minimal and non-unique set of molecules needed to explain the DFMS observations. While this set might not capture the full organic diversity, but rather its lower limit, it identifies many N- and NO-bearing COMs with reasonable certainty, while excluding others, potentially informing future observational campaigns, and hence contributes to the exploration of the origin and evolution of organic complexity in space. Among the key results is strong evidence for an abundant presence of heterocycles as well as substantial alkylation of both cyclic and acyclic species. These findings align well with reports on soluble organic matter in meteorites and asteroids and underpin once more the potential importance of such extraterrestrial organic material as a feedstock for terrestrial prebiotic chemistry.}


\maketitle

\section{Introduction}\label{sec:intro}
Heteroelements in organic molecules, i.e. oxygen (O), nitrogen (N), sulfur (S), and phosphorus (P), play crucial roles in the biochemistry of carbon-based life as they define the chemical functionality of a molecule or a molecular moiety. Thus, their study is of great interest already at the earliest stages of molecule formation in the interstellar medium \citep[ISM; e.g.][]{tielens2021}. However, while smaller complex organic molecules (COMs) such as methanol are common targets of remote spectroscopic observations of the ISM, the low volatility and relative abundances of larger COMs impose major obstacles towards their gas phase detection. As of mid-2021, more than 300 molecules have been detected \citep{mcguire2022} and current frontiers are being pushed further in the frameworks of large surveys of organic-rich environments such as the Taurus molecular cloud (TMC-1) \citep{mcguire2020,cernicharo2021a}. Nevertheless, we still lack a comprehensive view on the inventory of molecular complexity in pristine interstellar ices. Retrieved extraterrestrial samples have allowed us to work around the limitations of remote observations and provided insights into the chemical composition of Solar System organics. Carbonaceous chondrites and their parent bodies, the C-group asteroids that are thought to have formed in the outer Solar System before migrating inwards, harbour an extensive cache of extraterrestrial organics that are accessible to state-of-the-art analytical tools \citep{botta2002,schmitt-kopplin2010,alexander2017}. Samples returned from the various Solar System bodies have revealed valuable information: dust from comet 81P/Wild 2 returned by the National Aeronautics and Space Administration's (NASA's) Stardust mission \citep{sandford2008}, material from asteroid Ryugu returned by the Japan Aerospace Exploration Agency's (JAXA's) Hayabusa2 mission \citep{oba2023b,potiszil2023b,yabuta2023,schmitt-kopplin2023}, or material from asteroid Bennu returned by NASA's OSIRIS-REx mission \citep{glavin2025}. However, such analyses of extraterrestrial organics are affected by various types of processes causing chemical changes to the original material. Depending on the type of sample, these range from heating and shocks during entry into the Earth's atmosphere and impact during sampling procedures in space to terrestrial weathering prior to sample collection  and   analytical protocols (e.g. involving solvent extractions). In addition, asteroids, or rather their parent bodies, underwent different degrees of post-accretional processing, for instance by hydrothermal alteration, thermal metamorphism, or shocks. Effects of these alteration processes could be elucidated and constrained by comparative compositional studies of pristine Solar System matter like that in comets and asteroidal matter that has been processed to varying degrees.\\
It is therefore crucial to study pristine cometary organics, and the in situ data collected in the framework of the European Space Agency (ESA) Rosetta mission offer a unique opportunity to do so. The Rosetta spacecraft rendezvoused with its target, comet 67P/Churyumov-Gerasimenko (67P hereafter), in mid-2014 and accompanied it for 2 years on its orbit around the Sun. The Rosetta Orbiter Spectrometer for Ion and Neutral Analysis \citep[ROSINA;][]{balsiger2007} instrument suite and especially its high-resolution Double Focusing Mass Spectrometer (DFMS) revealed the chemical composition of cometary volatiles including COMs \citep{altwegg2016,rubin2019a,haenni2022,haenni2023,haenni2024} as they outgassed from the cometary nucleus and ejected particles. Before Rosetta, only a few N- and NO-bearing COMs had been identified from remote-sensing observations at millimetre wavelengths, as listed in Table \ref{tab:comets}, compiled from \citet{biver2019}. These molecules have also been identified in DFMS data as reviewed by \citet{altwegg2019}.

\begin{table*}[h!t]
   \caption[]{N-bearing molecules detected in comets by means of remote-sensing at millimetre wavelengths.$^{(a)}$}
   \centering 
   \begin{tabular}{lllll}
   \hline\hline
   Molecule                  & Constitution formula      & Mass (Da)  & Comment       & Identified in DFMS data$^{(b)}$  \\
   \hline
   Hydrogen cyanide          & HCN                       & 27         & detected      & yes$^{(c)}$      \\
   Hydrogen isocyanide       & HNC                       & 27         & detected      & no ref. spectrum \\
   Methylamine               & CH$_5$N                   & 31         & UL            & yes              \\
   Acetonitrile              & C$_2$H$_3$N               & 41         & detected      & yes$^{(d)}$      \\
   Cyanoacetylene            & C$_3$HN                   & 51         & detected      & yes              \\
   Acrylonitrile$^{(e)}$     & C$_3$H$_3$N               & 53         & UL            & yes              \\
   Ethylcyanide$^{(f)}$      & C$_3$H$_5$N               & 55         & UL            & yes              \\
   \hline
   Isocyanic acid            & HNCO                      & 43         & detected      & yes              \\
   Formamide                 & CH$_2$NO                  & 45         & detected      & yes              \\
   Glycine                   & C$_2$H$_5$NO$_2$          & 75         & UL            & yes$^{(g)}$      \\
   \hline
   \end{tabular}\label{tab:comets}
      \tablefoot{$^{(a)}$ The table is based on the review of \citet{biver2019}. For certain molecules not detected in the searches, upper limits (ULs) have been reported. $^{(b)}$ We indicate the identification of these molecules in DFMS data as reviewed by \citet{altwegg2019}. $^{(c)}$ For the isomer of HCN, i.e. HNC, no EI-MS reference spectrum is available. But the DFMS data is compatible with the fragmentation pattern of HCN, and it is unclear if and how much HNC contributes. $^{(d)}$ Acetonitrile and methylisocyanide cannot be distinguished in DFMS data as the differences in their fragmentation patterns are too little. $^{(e)}$ The IUPAC name is 2-propenenitrile. $^{(f)}$ The IUPAC name is propanenitrile. $^{(g)}$ Glycine has been reported by \citet{altwegg2016}.}
\end{table*}

\noindent An extraordinarily rich dataset was collected by the DFMS on 3 August 2015, while comet 67P was at 1.25 au and close to its perihelion at 1.24 au. \citet{haenni2022} reduced and investigated this dataset (for an overview, see their Fig.~1). They suggested that enhanced desorption from the ejected dust led to the detection of a plethora of volatile organics with molecular weights of up to at least 140 Da and an average sum formula of C$_1$H$_{1.56}$O$_{0.134}$N$_{0.046}$S$_{0.017}$. This average sum formula is similar to soluble organic matter from meteorites \citep{botta2002,alexander2017} and even to organics found in Saturn's ring rain \citep{miller2020} by Cassini's mass spectrometer. It could be indicative of a shared origin of that matter in the early Solar System. An Occam's razor-based deconvolution approach was used to attempt identification (or exclusion) of individual COMs as candidates present in the gas mixture of the cometary coma. After having discussed comet 67P's pure hydrocarbon budget in \citet{haenni2022} and the budget of O-bearing COMs in \citet{haenni2023}, the current work uses, also for the sake of comparability, the same approach to investigate the chemical nature and diversity of the N- and NO-bearing species of the general sum formula C$_n$H$_m$N and C$_n$H$_m$NO, respectively, where $n$ and $m$ are the stoichiometric coefficients of carbon and hydrogen. Signals of species with multiple N and O atoms are sparse and will be the topic of future work. Section \ref{sec:instr} revisits the key points of the instrument as well as the data reduction, and highlights the scientific value of the rationale behind the data interpretation approach. In Sect. \ref{sec:res} we present a list of observed C$_n$H$_m$N and C$_n$H$_m$NO parent sum formulas, and discuss potential candidate molecular structures that could reproduce the observations. Section \ref{sec:disc}  discusses the limitations of our interpretative approach, and compares our findings to the available literature on other well-studied reservoirs of Solar System organics, also highlighting  the (pre)biotic relevance of the chemical functionalities featured in our work. We conclude with a short summary and outlook in Sect. \ref{sec:sumconc}.

\section{Data evaluation}\label{sec:instr}
A detailed instrument description can be found in \citet{balsiger2007} The data acquisition, reduction, and evaluation procedures including error estimations for this specific dataset are explained in \citet{haenni2022,haenni2023}. In the following, we only give a brief summary thereof, focusing on the specificities of the fragmentation behaviour of N- and NO-bearing molecules.

\subsection{Method and instrumentation}\label{subsec:data-decon}
\begin{figure}
\centering
\includegraphics[width=8cm]{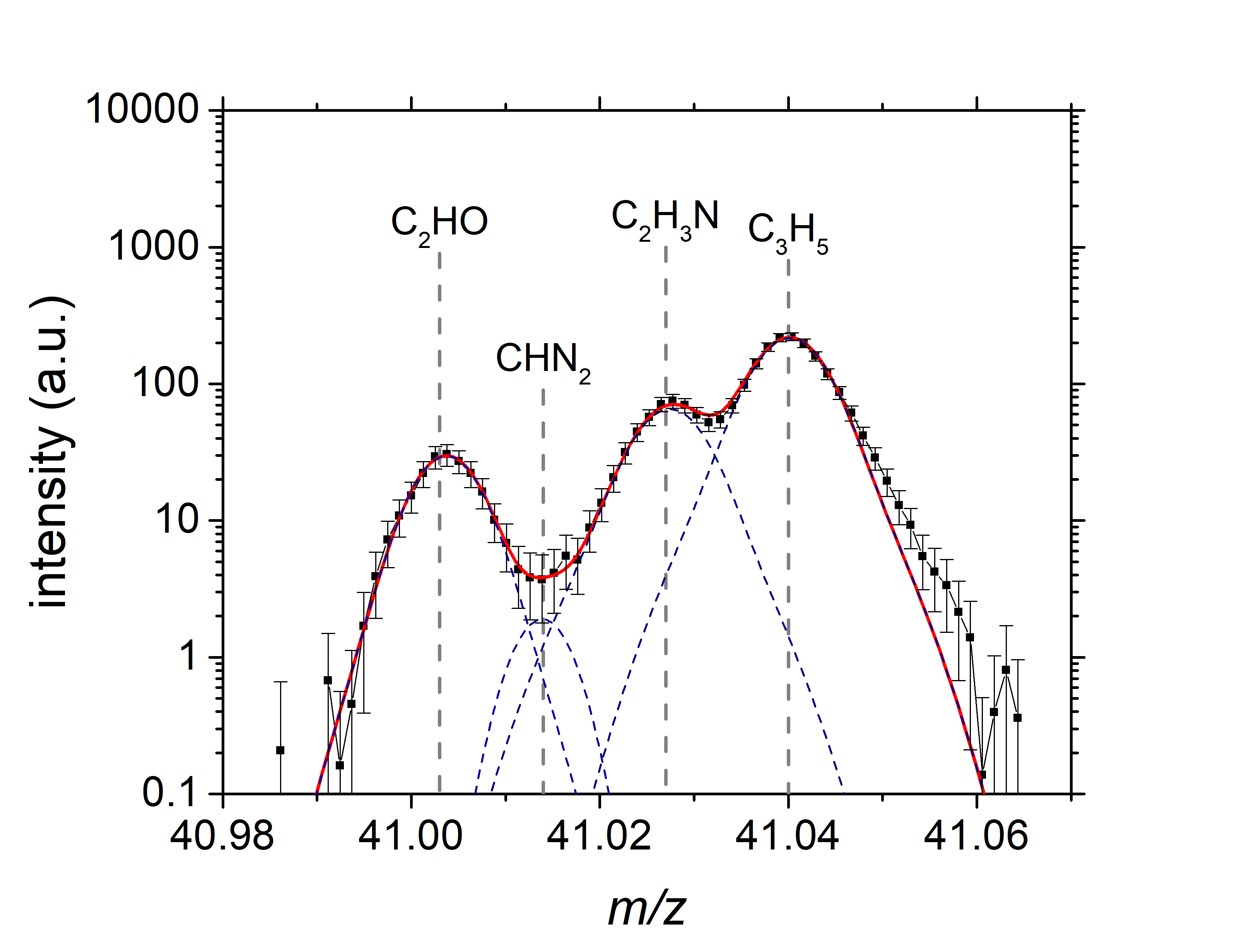}
   \caption{High-resolution mass spectrum collected on 3 August 2015 by the DFMS in 67P's coma around \textit{m/z} = 41 (black square markers). The peaks, associated with C$_3$H$_5$, C$_2$H$_3$N, CHN$_2$, and C$_2$HO at the exact positions indicated by the dashed grey lines, are normally fitted with double-Gaussian peak profile functions (dashed blue lines with the sum shown as a solid red line).}
      \label{fig:spec}
\end{figure}

A mass spectrometer allows  the characterization of an analyte, here the complex gas mixture of the cometary coma, in terms of the exact masses of the individual components. In the case of the DFMS, the gaseous species are ionized by electron ionization (EI) and the resulting cations are separated according to their mass-to-charge ratio (\textit{m/z}) by an electrostatic analyser and a subsequent curved permanent magnet. Eventually, the transmitted ions impinge on a microchannel plate-linear electron detector array assembly \citep{dekeyser2017} and the intensities of the signals are retrieved in arbitrary units (a. u.). As the detector yield and the ionization cross-sections are usually not known for COMs, the derivation of absolute local abundances (molecules/ccm) is not possible and we indicate the fragment sum relative to the common COM methanol. As water, commonly used to normalize cometary species, does not correlate well with species sublimating from dust, we have chosen to use methanol, frequently observed in the ISM, instead (for a more detailed explanation, see \citealt{haenni2023}).\\
A high-resolution DFMS mass spectrum covers just a narrow region of the mass range around the commanded integer mass. In order to cover a wider mass range, from \textit{m/z} = 13 to 140 (the upper end of the DFMS mass range at that time), as in the dataset investigated for this work, the DFMS scanned integer masses sequentially. Figure \ref{fig:spec} demonstrates that the DFMS mass resolution of \textit{m/$\Delta$m} = 3000 (at 1\% of the peak height at \textit{m/z} = 28; \citet{balsiger2007}) is sufficient to separate pure hydrocarbon species from heteroatom-bearing ones. In this work, we make use of DFMS' resolving power and attempt a full and untargeted analysis of the subset of C$_n$H$_m$N and C$_n$H$_m$NO signals collected on 3 August 2015 and shown in Fig.~1 in \citet{haenni2022}, panels (c) and (e). We have improved the drift correction, which results in slightly different intensities than those shown in Fig.~1 in \citet{haenni2022}. As detailed in \citet{haenni2022}, the drift in the observed signal intensities (due to changing local densities resulting from factors such as spacecraft or comet motion while stepping through the mass range) is corrected based on the interpolation of frequent water measurements. For masses below about \textit{m/z} = 40, the improved drift correction yielded slightly smaller intensities. Moreover, we do not consider signals in this evaluation that are only upper limits due to overlap with intense peaks usually from the pure hydrocarbon or O-bearing hydrocarbon on the same integer mass. For some masses, we also have improved our fits. General details about the DFMS data reduction routines have been reported previously \citep{leroy2015,rubin2019b,dekeyser2019,dekeyser2024}.\\
The major difficulty in interpreting high-resolution mass spectra of a complex gas mixture lies in EI-induced fragmentation. Parent species usually fragment into daughter species and the charge-retaining fragments are registered alongside (or instead of) the molecular ion (M) signal and together yield the so-called fragmentation pattern of the parent. For a gas mixture such as 67P's coma, the individual components as well as their EI-induced fragments are simultaneously present in the instrument's ionization chamber and as a result, a sum fragmentation pattern, i.e. a stack of individual fragmentation patterns, is measured. Information on how such a sum fragmentation pattern may be interpreted will be given in the following Subsect. \ref{subsec:challenge}.

\subsection{Sum fragmentation pattern deconvolution}\label{subsec:challenge}
To determine which chemical species are present at the time of data acquisition, the sum fragmentation pattern must be deconvolved. For the spectral deconvolution, reference mass spectra are needed, such as those collected in the extensive mass spectrometry database of the National Institute of Standards and Technology \citep[NIST;][]{NIST}. The ionization energy used to record the reference mass spectra archived in the NIST database normally is 70 or 75 eV while DFMS uses 45 eV. As it is impossible to calibrate every compound on the DFMS laboratory twin model, we have to rely on these standard mass spectra. It has been demonstrated for pure \citep{schuhmann2019b} and O-bearing hydrocarbons \citep{schuhmann2019a}, that the calibrated spectra deviate slightly from the standards in that relatively less of the fragments is observed. How these differences might systematically affect our evaluation will be discussed in Sect. \ref{sec:disc}. Moreover, the NIST reference mass spectra provide unit-resolution information only, which results in ambiguities for certain signals as to whether the heteroatom is retained in the fragment species or not. For example, a given molecule contains a N atom but also an ethyl group and makes a fragment on \textit{m/z} = 29. Now this fragment could either be due to the N-related functionality (i.e. CH$_3$N) or due to the ethyl group (i.e. C$_2$H$_5$) or even due to a mixture of both and high-resolution data are needed to clarify. Alternatively, sometimes deuterated or $^{13}$C isotopologues are measured to clarify the fragmentation mechanisms. For statistical reasons, if two heteroatoms are present in the parent, the instances of ambiguous fragments increase. Where no literature with high-resolution spectra of candidate molecules is available (and such literature on COMs is scarce), it is necessary to interpret those unit-resolution mass spectra. Therefore, we considered the rules and fundamentals of EI-MS-based structure determination as collected for instance in \citet{mikaia2022,mikaia2023} and summarized briefly below. In the following, '--' indicates chemical bonds (all valencies), whereas '-' indicates that a fragment with a certain mass was lost from the molecular ion M. For the sake of clarity, we omit indicating the positive charges resulting from the EI process.\\
--Stevenson's rule \citep{stevenson1951}: In the case of competing fragments, the fragment with the lowest ionization energy is usually the most abundant one and forms the base peak (i.e. the most intense peak in the fragmentation pattern of the molecule).\\
--McLafferty rearrangement in carbonyl-containing ions \citep{mclafferty1959}: The hydrogen atom on the $\gamma$-carbon (i.e. four atoms away from the carbonyl oxygen) is transferred in a 1,5-shift, which creates a distonic radical cation (i.e. charge and radical site are separated) while the 6-membered transition state is sterically favourable.\\
--Fragmentation of aliphatic amines (R--NH$_2$, where R is an aliphatic residue): Generally, the charge site is on the N atom, meaning that the N atom is retained while CH$_2$ groups (14 Da) are lost. Rarely, NH$_3$ (17 Da) is lost, leading to an \mbox{M-NH$_3$} signal. The largest aliphatic chain residue is lost preferentially because the resulting radical is more stable. In primary amines with an unbranched $\alpha$-carbon, the base peak is CH$_4$N on \mbox{\textit{m/z} = 30}, which corresponds to the resonance stabilized iminium ion. For methylated secondary amines, C$_2$H$_6$N on \textit{m/z} = 44 is the base peak as a result of $\beta$-cleavage. This peak increases in the homologous series of primary aliphatic amines with chain length. For dimethylated tertiary amines C$_3$H$_8$N on \textit{m/z} = 58, resulting from $\beta$-cleavage, is the base peak. If the amine group is located on an aromatic ring, usually M and M-H are strong. If the amine on the aromatic ring is a secondary amine (Ar--NH--R, where Ar is an aromatic ring), then Ar--NH--CH$_2$ on \textit{m/z} = 106, which results from $\beta$-cleavage at the R site, is resonance-stabilized and hence strong. If HCN (27 Da) is lost, then C$_5$H$_5$ on \textit{m/z} = 65 and C$_5$H$_6$ on \textit{m/z} = 66 are high.\\
--Fragmentation of N-heterocycles (R$_c$--N(H)--R$_c$, where R$_c$ is a cyclic structure): These species normally yield strong M and M-H signals because $\alpha$-cleavage, which leads to a ring-opening, can also entail loss of a hydrogen radical. Often, either M or M-H is the base peak.\\
--Fragmentation of nitriles (R--CN): Nitriles tend to produce M-H after losing an $\alpha$-hydrogen atom but no M. Hence, they are difficult to identify. A prominent and frequent base peak at resonance-stabilized NH-C-CH$_2$ on \textit{m/z} = 41 is the result of the McLafferty rearrangement in compounds whose $\alpha$-carbon atom is not branched. Straight-chain nitriles containing seven or more carbon atoms lead to a characteristic signal on \textit{m/z} = 97, which corresponds to a 6-membered ring of carbon atoms with an NH-substituent (C$_6$H$_11$N).\\
--Fragmentation of carboxamides (R--C(O)--NR'R'', where R' and R'' are two more residues): CH$_2$NO on \textit{m/z} = 44 is a common base peak of primary amides if the abstractable hydrogen for the McLafferty rearrangement is missing. Primary amides also produce a 5-membered cyclic amide fragment on \textit{m/z} = 86. If there is an abstractable hydrogen, then C$_2$H$_5$NO on \textit{m/z} = 59 is the base peak for primary, secondary, and tertiary amides. CNH$_4$N on \textit{m/z} = 30 (as for amines) is obtained, if the alkyl groups on the nitrogen are longer than two carbon atoms. Aromatic amides with alkyl substituents lead to Ar-CO (i.e. M-NR$_2$). Subsequent loss of CO leads to C$_6$H$_5$ on \textit{m/z} = 77.\\
\noindent The key points summarized here capture the typical fragmentation behavior of the most common groups of N-bearing molecules and are a guide to interpreting the unit resolution reference spectra available from NIST \citep{NIST}.\\
Generally, a specific molecule in a mass spectrum of a complex mixture like that of  the cometary coma can be identified unambiguously if two conditions are met: First, if it has a detectable M signal plus fragment peaks that are compatible with the observed pattern. Note that fragment signals alone are ambiguous under such conditions as many species could yield the same fragments. Second, if no alternative structural isomer exists or all but one can be ruled out. The latter is only the case for small molecules as the number of structural isomers grows with the number of atoms in the molecule, resulting in a combinatorial explosion. Mass spectrometry, by default, cannot distinguish between structural isomers as they have the same exact mass, unless their fragmentation patterns differ significantly. One example, where the distinction of two structural isomers (dimethylsulfide and ethanethiol) is possible, has been presented by \citet{haenni2024}. On the other hand, if a given molecule shows a signal (parent or fragment) in its fragmentation pattern that is not present in the observed data, then this is an exclusion criterion. In that way, certain molecules can be clearly excluded while it may still not be possible to pin down one specific candidate molecule for an observed parent sum formula.\\
In summary, to derive the true composition of the cometary neutral coma based on the sum fragmentation pattern observed by the DFMS on 3 August 2015 and presented in \citet{haenni2022}, a spectral deconvolution into individual components is necessary. In the following paragraph, we argue that strictly mathematical deconvolution methods are too limited and explain which approach we pursued instead.

\subsection{Data interpretation strategy}\label{subsec:strategy}
\noindent Mass spectra deconvolution often employs mathematical approaches such as residual minimization. As a consequence of the uncertainties mentioned above as well as in \citet{haenni2022} and \citet{haenni2023}, the mathematical problem is ill-conditioned and minimizing residuals is equivalent to finding a global minimum in a flat subspace. We instead adopt a discussion-driven approach to explore the complex data, incorporating additional astrochemical constraints. On the one hand, we take into account observational reports from studies of relevant astrophysical environments such as other comets or the ISM. On the other hand, we assume a bottom-up chemistry, where (organic) molecules are built up from atomic building blocks in the first place, which should lead to decreasing abundance with increasing size of the molecule, a trend reflecting in the DFMS data overviewed in Fig. 1 in \citet{haenni2022}. Bottom-up chemical reaction networks are preferentially used to explain COMs in the ISM but are debated for larger carbon-based species such as polycyclic aromatic hydrocarbons \citep{lee2019}. The credo of a bottom-up logic is that high abundances of a complex species do not make sense if simpler species in the bottom-up formational pathway have much lower abundances. We also favour bottom-up chemistry intermediates where compatible with our data. Guided by Occam's razor \citep{clauberg1654} and in combination with the above constraints, we arrive at a non-unique set of cometary COM candidates that can explain the DFMS data. This set is not only non-unique but it almost certainly underestimates the true chemical complexity for COMs with many isomers. Nevertheless, it sets a lower limit to the chemical diversity and complexity. More important though, is a thorough discussion of every observed parent sum formula and potential candidate molecular structures as far as the available reference data allows. Therefore, we do not compile a list of the selected candidates but rather refer to   our Sect. \ref{sec:res}. We assign each selected candidate molecule with a level-of-confidence indicator (LCI) that could also be viewed as a degeneracy indicator: 1 = maybe present, 2 = likely present, 3 = with great certainty present. Notably, our LCIs are determined with respect to the available reference data. This means that a certain candidate is assigned with LCI = 3, if it is the only isomer with reference data in NIST and is in reasonable agreement with our data. It still may be possible, though, that other isomers, for which no reference fragmentation data exists, would be an even better match. Specific caveats and biases resulting from our approach and the application of Occam's razor to deconvolve DFMS mass spectra will be discussed thoroughly in Sect. \ref{sec:disc}. We  also address some limitations of the method of EI-MS itself to capture organic complexity.

\section{Results}\label{sec:res}
\begin{figure*}
  \centering
  \includegraphics[width=\textwidth]{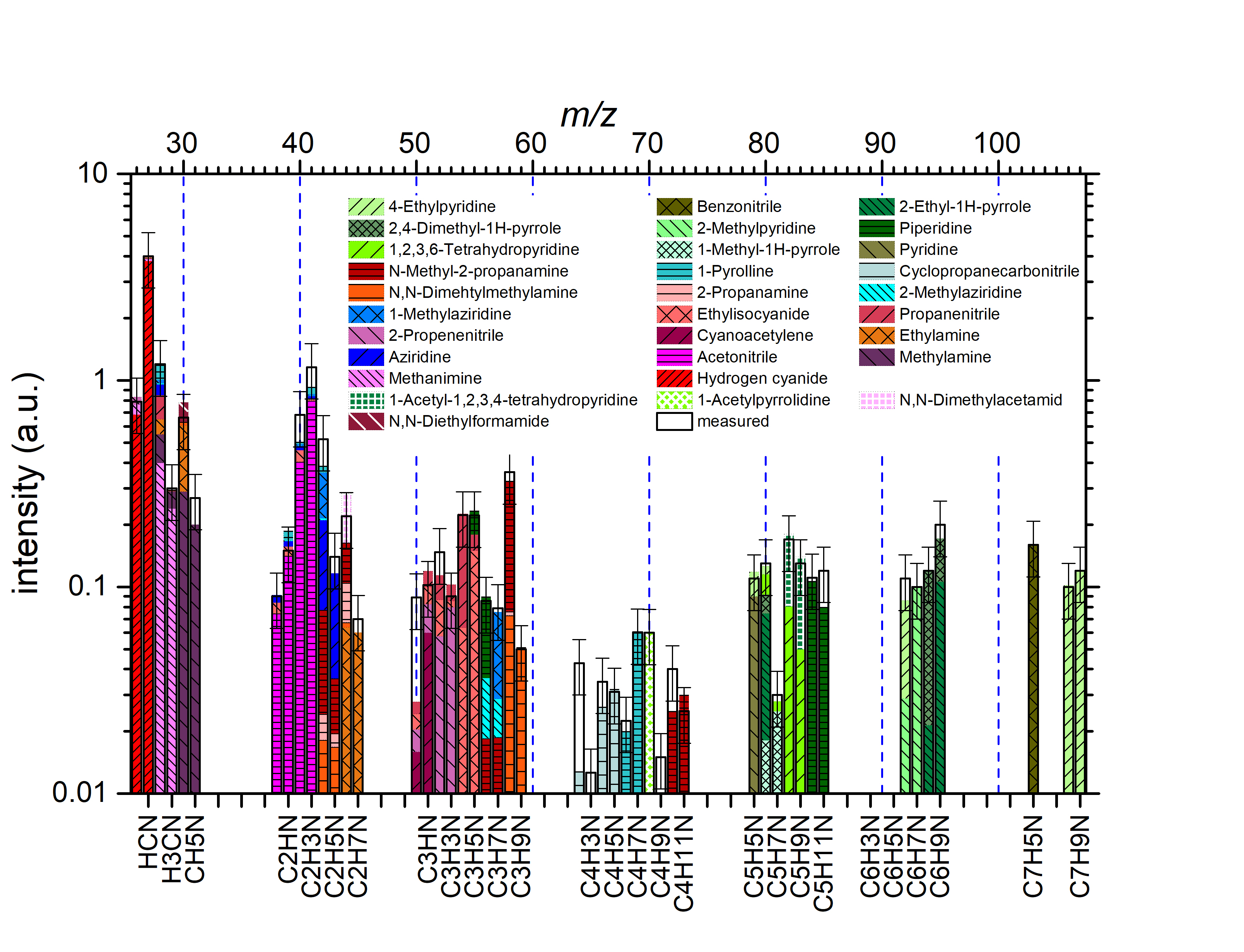}
     \caption{Occam's razor-based deconvolution result of the subset of signals associated with C$_n$H$_m$N parent and fragment species detected by the DFMS on 3 August 2015. The measured signals (measured) are given in arbitrary units (a. u.) with 30~\% error margins. The different contributions of the chosen candidate molecules are colour-coded according to their molecular structure -- in shades of red and purple for chain-based molecules and in shades of green and blue for cyclic molecules.}
        \label{fig:CHN}
\end{figure*}

Figs. \ref{fig:CHN} and \ref{fig:CHNO} show our non-unique Occam's razor-based deconvolution results for the C$_m$H$_n$N and C$_m$H$_n$NO sum spectra observed by Rosetta's DFMS on 3 August 2015, accounting for all relevant signals. Our analysis assumes these two subsets of data to be largely independent, which seems justified for relative abundance reasons discussed in detail previously \citep{haenni2022,haenni2023}. However, certain interdependences are expected as many C$_m$H$_n$N and C$_m$H$_n$NO parents have fragments of other classes than C$_m$H$_n$N and C$_m$H$_n$NO, namely when heteroatom-bearing fragments are lost. For instance, C$_m$H$_n$NO parents often yield C$_m$H$_n$N, C$_m$H$_n$O, or C$_m$H$_n$ type fragments (and not just C$_m$H$_n$NO type fragments) and we take such interdependences into consideration and discuss them where needed. We applied Occam's razor starting from the smallest molecules with the least number of isomers and hence the least ambiguity and this is also how we present the results here. For every parent sum formula, we discuss potential candidate molecular structures with available reference spectra and assign the most suitable one with an LCI as described above. Notably, fully saturated species are always parents as such fragments are rare and if produced then normally lost as neutral molecules. Generally, neutral parent molecules of the C$_m$H$_n$N and C$_m$H$_n$NO type must have an odd nominal mass, even mass representatives are always fragments. From the candidate selected to represent a parent, we use the sum over all observed ions from the reference spectrum to derive the fragment sum relative to methanol (FSRM; in \%). We have chosen methanol because the investigated data captures outgassing predominantly from grains rather than from the nominal, water-dominated gas coma and because this simple COM is a common reference species in observations of various astrophysical environments including the ISM \citep{haenni2023}.

\subsection{C$_n$H$_m$N species}\label{subsec:CHN}

--HCN (\textit{m/z} = 27): Hydrogen cyanide (LCI = 3; FSRM = 81~\%) is identified unambiguously via M and M-H. However, a contribution of its isomer HNC cannot be excluded as the isomer's fragmentation pattern is unknown. Both isomers have been reported to be present in cometary outgasing from ground-based observations of various comets. For comet Hyakutake, an HNC/HCN ratio of $\approx0.06$, which is similar to that observed in quiescent molecular clouds, was presented as evidence for ISM inheritance of cometary ices \citep{irvine1996}. However, the ratio of isomers seems to vary with cometocentric and heliocentric distances, suggesting multiple sources or chemical processes at play \citep{biver1997,rodgers1998,agundez2014}. The debate is ongoing.\\
--CH$_3$N (\textit{m/z} = 29): Methanimine (LCI = 2; FSRM = 5.1~\%) is identified via M and M-H under assumption of a fragmentation analog to formaldehyde. The fragmentation pattern of methanimine has not been measured to the best of our knowledge.\\
--CH$_5$N (\textit{m/z} = 31): Methylamine (LCI = 3; FSRM = 17~\%) is identified unambiguously via M, M-H, M-2H, and M-3H. Notably, methylamine does not have any structural isomers which is why we can exactly determine its contribution to the iminium ion signal on \textit{m/z} = 30.\\
--C$_2$H$_3$N (\textit{m/z} = 41): Acetonitrile (LCI = 3; FSRM = 14~\%) is identified via M and M-H. But the EI-MS fragmentation pattern of acetonitrile is indistinguishable from that of its structural isomer, methylisocyanide. We assign all intensity to acetonitrile but it could be assigned as well to methylisocyanide or any mixture of the two.\\
--C$_2$H$_5$N (\textit{m/z} = 43): Aziridine (LCI = 3; FSRM = 8.3~\%) is identified based on the M-H/M ratio and the M-CH$_3$ fragment. Ethanimine might be a reasonable alternative if it fragments similarly to methanimine/formaldehyde, which produces quite strong M and M-H signals. Unfortunately, no EI-MS reference data are available. According to \citet{mcguire2022}, aziridine was detected tentatively in the ISM while ethanimine was conclusively identified (also see discussion below).\\
--C$_2$H$_7$N (\textit{m/z} = 45): Ethylamine (LCI = 2; FSRM = 13~\%) is identified via M, M-H, and M-CH$_3$. Dimethylamine, the only isomer of ethylamine, is a possible alternative candidate. Dimethylamine would almost fully account for the C$_2$H$_6$N fragment on \textit{m/z} = 44 via its M-H fragment, just as ethylamine fully accounts within error margins for the residual CH$_4$N intensity on \textit{m/z} = 30, i.e. the intensity that is not explained already by methylamine. Since both these fragments are characteristic for the whole class of amines, the abundance of amines with 3 or more C atoms is strongly limited, no matter which isomer is chosen for C$_2$H$_7$N. Ethylamine but not dimethylamine has been detected in acid-hydrolyzed samples returned by NASA's Stardust mission from comet Wild2 \citep{glavin2008} and in hot water extracts of Ryugu samples \citep{naraoka2023} (for more details, see Sect. \ref{subsec:context}) and \citet{altwegg2016} have reported the presence of ethylamine in comet 67P. In the ISM, a tentative detection of ethylamine towards TMC-1 \citep{zeng2021} and a non-detection of dimethylamine towards the same source \citep{mueller2023} unfortunately do not constrain the relative abundance ratio of the two isomers much. However, in experimental work on irradiated ices, ethylamine shows a better compatibility with the data than its isomer \citep{foerstel2017}. Therefore, we choose ethylamine for our solution while acknowledging that this choice influences the selection especially of higher-mass amines.\\
--C$_3$HN (\textit{m/z} = 51): Cyanoacetylene (LCI = 2; FSRM = 1.6~\%) is identified via M and M-H. However, the M-H/M ratio does not fit perfectly and it cannot be excluded that the structural isomer isocyanoacetylene would produce a better fit. Unfortunately, no fragmentation data of isocyanoacetylene are available. Both these isomers, as well as the isomer of the structure HNCCC, have been reported in the ISM \citep{mcguire2022,cernicharo2020}. An alternative explanation for the underpopulated M-H signal on \textit{m/z} = 50 could be the presence of the C$_3$N radical (cyanoethinyl; see further discussions in Subsect. \ref{subsec:unexplained}).\\
--C$_3$H$_3$N (\textit{m/z} = 53): 2-Propenenitrile (LCI = 3; FSRM = 4.3~\%) is identified via M, M-H, and M-2H. However, the M-H/M ratio does not fit well, which might indicate the (additional) contribution of other isomers. But no fragmentation data are available for any of them.\\
--C$_3$H$_5$N (\textit{m/z} = 55): Ethylisocyanide (LCI = 2; FSRM = 15~\%) is identified via M and M-H as the most likely candidate. However, it cannot be distiguished well from N-methyleneethenamine, which has a very similar fragmentation pattern. Propanenitrile or propargylamine yield substantially more M-H and a contribution is needed to explain the observed M-H signal (LCI = 2; FSRM = 9.2~\%; considering the fragmentation pattern of propanenitrile). The dehydrogenated four-membered ring, 1-azetine, as well as vinylimine do not have reference fragmentation data on NIST.\\
--C$_3$H$_7$N (\textit{m/z} = 57): A combination of 1-methylaziridine (LCI = 3; FSRM = 4.4~\%) and 2-methylaziridine (LCI = 3; FSRM = 1.1~\%) best explains the observed intensity via M, M-H, and M-CH$_3$. Cyclopropylamine and 2-propen-1-amine do have compatible M/M-H ratios, but they fail to explain the residual intensity on \textit{m/z} = 42, which is assigned to a fragment after loss of a methyl-group (especially strong in the fragmentation pattern of 1-methylaziridine). Azetidine neither explains the intensity on \textit{m/z} = 42 nor shows a compatible M/M-H ratio which strongly constrains its presence.\\
--C$_3$H$_9$N (\textit{m/z} = 59): A combination of N,N-dimethylmethylamine (LCI = 2; FSRM = 3.4~\%) and 2-propanamine (LCI = 2; FSRM = 1.1~\%) suitably explains the observations. The former mainly yields M and M-H, while the latter yields M-H and M-CH$_3$. One alone cannot reproduce the M/M-H ratio. Propylamine is not favourable as the observed signal on CH$_4$N, the typical primary amine EI-fragment on \textit{m/z} = 30, is already well explained and a substantial abundance of larger primary amines is hence unlikely.\\
--C$_4$H$_5$N (\textit{m/z} = 67): Cyclopropanecarbonitrile (LCI = 1; FSRM = 5.9~\%) is the most likely candidate, identified via M, M-H, M-2H, and M-3H, where M-H is stronger than M, although several other isomeric nitriles produce a similar M/M-H ratio. This group of signals could also contain some pyrrol, which, however, has a strong M but no M-H and is hence less favourable according to Occam's razor. The unexplained intensity at M-2H (C$_4$H$_3$N) and especially at M-3H (C$_4$H$_2$N) might indicate a missing parent with the sum formula C$_4$H$_2$N. Further details will be discussed in Subsect. \ref{subsec:unexplained} later on.\\
--C$_4$H$_7$N (\textit{m/z} = 69): 1-Pyrroline (LCI = 2; FSRM = 1.3~\%) is the best suitable candidate. In this case, the reference mass spectrum has been taken from \citet{spectrabase} as it is not available from NIST. 1-Ethenylaziridine, reproduces the M/M-H ratio and additional fragments reasonably well. Therefore, it is a viable alternative.\\
--C$_4$H$_9$N (\textit{m/z} = 71): Pyrrolidine yields relevant M and M-H signals and a base peak on \textit{m/z} = 43, which corresponds to C$_2$H$_5$N and forms through cleavage of one C--C bond adjacent to the N atom and subsequent ethylene elimination \citep{spiteller1967}. However, C$_4$H$_8$N (M-H) cannot be sufficiently explained, calling for additional contributions. Instead of a combination from multiple parents, Occam's razor suggests consideration of 1-acetylpyridine (a C$_n$H$_m$NO species, which will be discussed below). 1-Acetylpyridine sufficiently covers C$_4$H$_8$N via its M-Ac (Ac = acetyl group) fragment. But a small contribution of pyrrolidine or other C$_4$H$_9$N isomers that yield a relevant M signal is likely.\\
--C$_4$H$_{11}$N (\textit{m/z} = 73): N-Methyl-2-propanamine (LCI = 2; FSRM = 3.8~\%) is a good candidate. N-Ethyl- or N,N-dimethylethanamine are less likely because the respective ratios of M, M-H and M-CH$_3$ are less compatible with the observations. Minor contributions cannot be ruled out, though.\\
--C$_5$H$_5$N (\textit{m/z} = 79): Pyridine (LCI = 3; FSRM = 3.5~\%) is identified via M and absence of M-H. 2,4-Pentadienenitrile, the only isomer with an available reference mass spectrum, has a considerably less prominent M signal and is hence less likely. The M-H signal should be about 12\% with respect to the M signal for pyridine and even more for 2,4-pentadienenitrile. However, as C$_5$H$_4$N is hidden under the strong C$_6$H$_6$ peak on \textit{m/z} = 78, only an upper limit of roughly 15\% (with respect to M) can be estimated, compatible with pyridine. Both species yield a pronounced fragment on \textit{m/z} = 52, which likely corresponds to M-HCN and hence to C$_4$H$_4$. Indeed, C$_4$H$_4$ is still underpopulated as can be seen in supplementary Fig.~1 in \citet{haenni2022}.\\
--C$_5$H$_7$N (\textit{m/z} = 81): 1-Methyl-1H-pyrrole (LCI = 1; FSRM = 1.4~\%) is selected as it is compatible with the observations in terms of M and M-H, given an additional contribution of 2-ethyl-1H-pyrrole to C$_5$H$_6$N via its M-CH$_3$ fragment (see below). Alternatively, there are molecules that fit the observed M/M-H ratio better, for example pyrroles with the methyl group on positions 2 and 3 instead of 1 (N). Isomers with nitrile functionality expose  M/M-H ratios similar to that of  the N-substituted pyrrol but, in addition, a clear signal on \textit{m/z} = 54. Probably HCN is lost, resulting in a C$_4$H$_6$ fragment on \textit{m/z} = 54. However, in supplementary Fig.~1 in \citet{haenni2022} C$_4$H$_6$ is already slightly overpopulated be beyond the error margin, making nitriles less likely candidates. For 1,4-dihydropyridine no fragmentation data are available but it may be a good candidate, given the fact that tetrahydropyridine is a good candidate as well and the presence of molecules with different degrees of hydrogenation has been previously reported \citep{haenni2022,haenni2023}.\\
--C$_5$H$_9$N (\textit{m/z} = 83): 1,2,3,6-Tetrahydropyridine (LCI = 2; FSRM = 9.2~\%) is identified via M and M-H. The double bond might not be on the N atom as the resulting fragmentation pattern, with an absent M-H fragment, is less compatible. Most isomers, such as isocyano, nitrile, or amine species, do not yield relevant M. Only the ternary amine N,N-dimethyl-2-propyn-1-amine would be a possible alternative.\\
--C$_5$H$_{11}$N (\textit{m/z} = 85): Piperidine (LCI = 2; FSRM = 6.0~\%) explains the observed M and M-H signals well, while its other fragments are reasonably compatible with the observations. 1-Methylpyrrolidine would show strong signals on \textit{m/z} = 57 (C$_4$H$_9$ from N--C-bond alpha-cleavage) and \textit{m/z} = 42 (C$_3$H$_6$ from ring-cleavage). Both these signals are already overpopulated in supplementary Fig.~1 in \citet{haenni2022}, which renders the N-substituted pyrrolidine incompatible. 2-Methylpyrrolidine as well as N-ethyl-2-propen-1-amine yield strong M-CH$_3$ signals on \textit{m/z} = 70 that cannot be accommodated in the observed data. While secondary and tertiary amines such as N-allyl-N,N-dimethylamine show favourable M/M-H ratios, they overpopulate the observed C$_3$H$_8$N signal on \textit{m/z} = 57. A contribution of 3-methylpyrrolidine cannot be ruled out.\\
--C$_6$H$_7$N (\textit{m/z} = 93): 2-Methylpyridine (LCI = 1; FSRM = 5.9~\%) is selected as it is compatible with the data in terms of M, M-H, and M-CH$_3$, considering that the M-CH$_3$ fragment of the selected C$_7$H$_9$N parent (4-ethyl-pyridine) is coincident with M-H of 2-methylpyridine. However, other molecules such as N-2-propynyl-2-propyn-1-amine (also known as dipropargylamine), yielding larger M-H signals, would lead to a better fit if for C$_7$H$_9$N (see below) a parent was selected that does not yield M-CH$_3$. Furthermore, the position of the methyl group is badly constrained from the fragmentation pattern alone. Aniline, also known as phenylamine, does not fit as well due to its relatively small M-H signal. A contribution cannot be ruled out, though.\\
--C$_6$H$_9$N (\textit{m/z} = 95): A combination of 2-ethyl-1H-pyrrole (LCI = 2; FSRM = 0.7~\%) and 2,4-dimethyl-1H-pyrrole (LCI = 2; FSRM = 3.5~\%) explains well the observations in terms of the M, M-H, and M-CH$_3$ signals, the latter being produced only from the ethyl-substituted pyrrole. The ethyl-substituted pyrrole yields more M-H while the dimethyl-substituted one yields more M. From the isomers with reference mass spectra on NIST, only the 1H-pyrrole derivatives produce reasonably strong M signals and hence might contribute while all other isomers do not have strong M and, hence, are not suitable candidates.\\
--C$_7$H$_5$N (\textit{m/z} = 103): Benzonitrile (LCI = 2; FSRM = 6.0~\%) or 2-ethynylpyridine are both compatible with the observations and cannot be distinguished based on their fragmentation patterns. \citet{haenni2022} have previously suggested the presence of benzonitrile and written that other isomers are less likely. However, within error margins, the two said isomers are indistinguishable. No reference data are available for other structural isomers. Benzonitrile was detected towards TMC-1 \citep{mcguire2018a}.\\
--C$_7$H$_9$N (\textit{m/z} = 107): 4-Ethylpyridine (LCI = 2; FSRM = 8.6~\%) is identified via M, M-H, and M-CH$_3$. The M/M-H ratio fits perfectly. Alternatively, a secondary amine on an aromatic ring (Ar) could lead to Ar--NH--CH$_2$ on \textit{m/z} = 106. Therefore, \citet{haenni2022} have suggested the presence of benzylamine. However, the M/M-H ratio of benzylamine is less suitable than that of 4-ethylpyridine and other alkylated pyridine derivatives. Only the isomers with more M-H than M, or no M and M-H at all, can be deemed incompatible with Occam's razor. Benzylamine is a possible minor contributor but it is more likely that the N atom is in the ring of a 6-membered heterocycle. C$_7$H$_9$N is the largest N-bearing species observed in this dataset.

\subsection{C$_n$H$_m$NO species}\label{subsec:CHNO}

\begin{figure*}
  \centering
  \includegraphics[width=\textwidth]{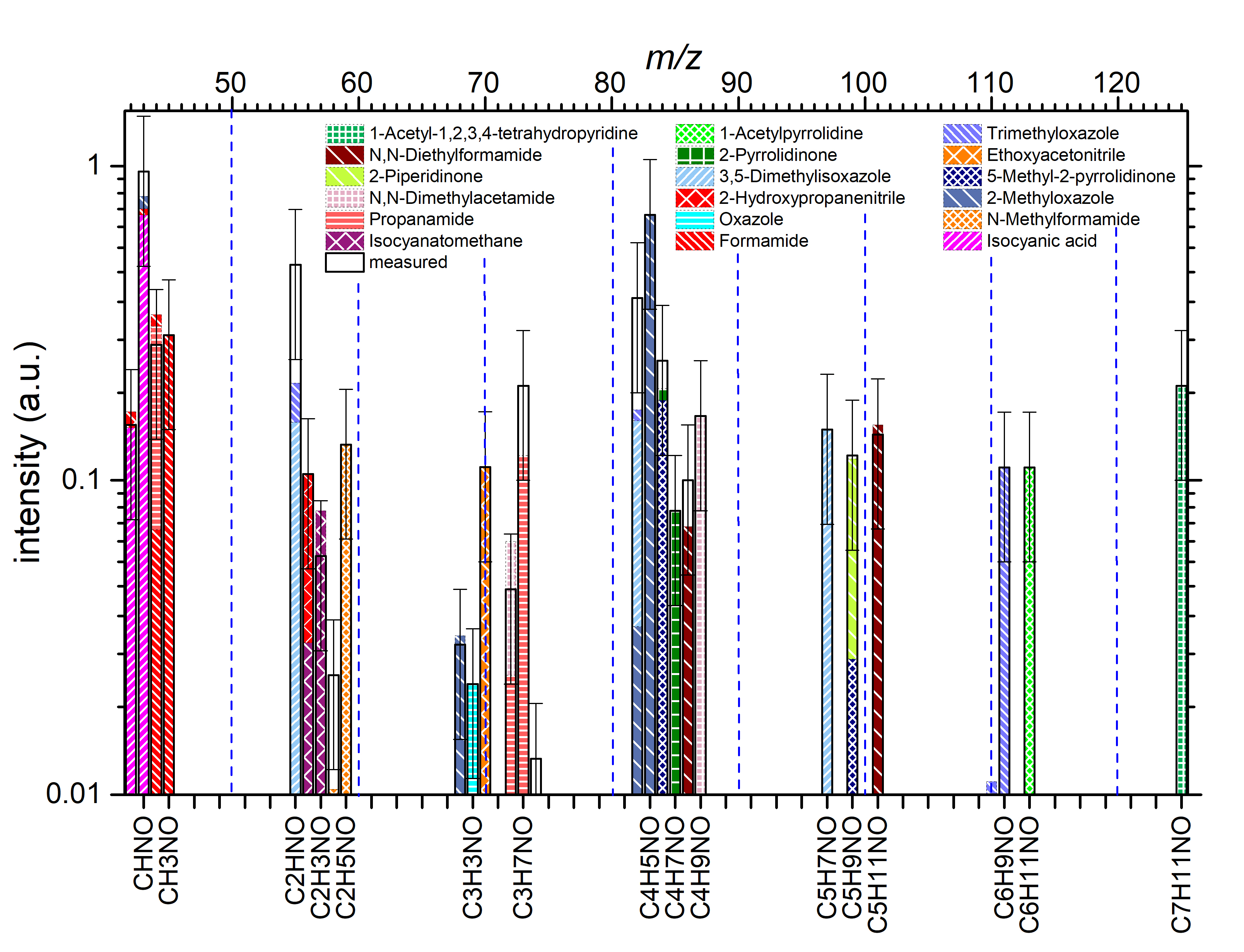}
     \caption{Occam's razor-based deconvolution result of the subset of signals associated with C$_n$H$_m$NO parent and fragment species detected by the DFMS on 3 August 2015. The measured signals (measured) are given in arbitrary units (a. u.) with 30\% error margins. The different contributions of the chosen candidate molecules are colour-coded according to their molecular structure -- in shades of red and purple for chain-based molecules and in shades of green and blue for cyclic molecules.}
        \label{fig:CHNO}
\end{figure*}

--CHNO (\textit{m/z} = 43): Isocyanic acid (LCI = 3; FSRM = 14~\%) is identified via M and M-H. The reference mass spectrum, which is not available from NIST, was taken from \citet{hand1971}. For its structural isomer, cyanic acid, the PubChem database \citep{pubchem} reports the liquid chromatography-mass spectrometry (LC-MS) spectrum, which was recorded using electro-spray ionization (ESI). Hence, this spectrum contains an M+H signal on \textit{m/z} = 44. Nevertheless, the LC-MS spectrum indicates that the M/M-H fragment ratio of cyanic acid might be less compatible with our observations. For other structural isomers no reference data are available.\\
--CH$_3$NO (\textit{m/z} = 45): Formamide (LCI = 3; FSRM = 12~\%) is unambiguously identified via M and M-H. The CH$_4$N fragment (iminium ion) on \textit{m/z} = 30 in the reference spectrum of the structural isomer nitrosomethane cannot be accommodated. For the third isomer, formadehyde oxime, no reference data are available. While formamide is a common molecule in the ISM, none of its isomers has been reported to date \citep{mcguire2022}.\\
--C$_2$H$_3$NO (\textit{m/z} = 57): Isocyanatomethane (LCI = 3; FSRM = 2.9~\%) is identified via M and M-H. It is likely that the fragments of isocyanatopropane contribute to the \textit{m/z} = 57 and 56 signals (see further discussion below) as loss of the carbonyl function is probably negligible compared to loss of the terminal ethyl group. This makes the presence of hydroxyacetonitrile unnecessary according to Occam's razor. For other isomers, no reference mass spectra are available.\\
--C$_2$H$_5$NO (\textit{m/z} = 59): Acetamide (LCI = 2; FSRM = 4.6~\%) is identified via a strong M, an absent M-H, and a very strong M-CH$_3$ signal. M-NH$_2$ and M-NH$_3$ fragments on \textit{m/z} = 43 and 42, respectively, do not appear in Fig. \ref{fig:CHNO} as the N is lost. The other isomers, for which reference data are available, are less likely, except N-methylformamide, which potentially contributes to the observed signals and also has been reported to be present in the ISM \citep{belloche2019}.\\
--C$_3$H$_3$NO (\textit{m/z} = 69): Oxazole	(LCI = 2; FSRM = 1.3~\%) is identified via strong M and negligible M-H. An alternative is propiolamide, which also yields a strong M but no M-H signal. However, this molecule's base peak on \textit{m/z} = 53 is ambiguous and no high-resolution data are available as far as we know. Thus, we cannot make clear statements in this case. Isoxazol, in contrast, can be ruled out based on its strong M-H fragment. However, C$_3$H$_2$NO, corresponding to M-H, is observed, but it is already explained by the M-CH$_3$ fragment of 2-methyloxazole. We  further discuss such entanglements below.\\
--C$_3$H$_7$NO (\textit{m/z} = 73): Propanamide (LCI = 2; FSRM = 9.9~\%) is chosen because its M, M-H, and M-CH$_2$CH$_3$ signals fit well. N,N-Dimethylformamide also fits well based on its M and M-CH$_3$ signals. According to \citet{li2016}, its strong fragment on \textit{m/z} = 44 corresponds to M-CHO (C$_2$H$_6$N) and the one on \textit{m/z} = 42 to M-CH$_3$O (C$_2$H$_4$N). However, especially the observed C$_2$H$_6$N signal in Fig. \ref{fig:CHN} is already overpopulated, which limits the contribution of N-dimethylformamide. The fragments of the other isomers with reference mass spectra do not fit well and they are thus unlikely candidates.\\
--C$_4$H$_5$NO (\textit{m/z} = 83): 2-Methyloxazole (LCI = 2; FSRM = 21~\%) is identified via M, M-H, and M-CH$_3$. The reference spectra for several oxazoles can be retrieved from spectrabase \citep{spectrabase} as on NIST only isoxazole derivatives are available and are less compatible with Occam's razor as they only yield M signals. The position of the methyl group, however, is not well defined as the reference spectra are similar. The high FSRM might indicate that several different oxazole species are simultaneously present as the effective abundance of specific COMs should decrease with increasing number of atoms. Nevertheless, we abstain from introducing more isomers with  the methyl group in a different position, for example, and instead just note that if the FSRM was distributed on three parent molecules, it would better align with the expected abundance decrease.\\
--C$_4$H$_7$NO (\textit{m/z} = 85): 2-Pyrrolidinone (LCI = 2; FSRM = 6.6~\%) yields M and M-H signals. High-resolution data are available from \citet{duffield1964}. Further isomers with reference spectra can be ruled out based on their fragmentation patterns mismatching the observations. However, most of the observed M-H intensity is attributed to the M-CH$_3$ fragment of the C$_5$H$_9$NO parent 5-methyl-2-pyrrolidinone (see below). This creates an entanglement and, hence, we do not assign LCI = 3 in this case.\\
--C$_4$H$_9$NO (\textit{m/z} = 87): N,N-Dimethylacetamide (LCI = 1; FSRM = 12~\%) is identified via its strong M and M-CH$_3$ signals. Moreover, the strong signal on \textit{m/z} = 44 corresponds to M-Ac and hence to C$_2$H$_6$N. This fragment appears in Fig. \ref{fig:CHN}. N-Methylpropanamide or O-methyloxime-2-propanone also yield M but no M-H and hence could be viable alternative candidates. N-Ethyl-acetamide is not likely present due to a strong iminium ion signal which cannot be accommodated. Morpholine and 2-butanoneoxime are possible candidates (they make M and 30-40\% M-H), if N,N-diethylformamide (the parent selected for C$_5$H$_{11}$NO) is replaced with a molecule that does not yield M-CH$_3$. This interference of M-CH$_3$ fragments of heavier species with M-H fragments is observed in various places and makes the interpretation of the mass spectrum more ambiguous (see further discussion below). Other isomers with available reference spectra do not yield relevant M signals and hence are not likely candidates.\\
--C$_5$H$_7$NO (\textit{m/z} = 97): 3,5-Dimethylisoxazole (LCI = 2; FSRM = 11~\%) yields compatible M and M-CH$_3$ signals and no M-H signal. All other isomers with reference data are less compatible as they either produce too much M-H or too little M-CH$_3$. Notably, the observed intensity on C$_4$H$_4$NO, i.e. the M-CH$_3$ fragment of the selected candidate 3,5-dimethylisoxazole, is still underexplained but we did not find a more suitable combination of molecules. Contributions of the oxazole derivatives among the isomers cannot be ruled out.\\
--C$_5$H$_9$NO (\textit{m/z} = 99): 2-Piperidinone (LCI = 2; FSRM = 8.3~\%) is compatible with its strong M and negligible M-H signals but its FSRM would be too high. We therefore introduce an additional parent: 5-Methyl-2-pyrrolidinone (LCI = 2; FSRM = 5.4~\%). This molecule can appropriately explain the residual M and M-CH$_3$ signals. All isomers with reference spectra that do not yield a substantial M signal or have a relevant M-H fragment are not compatible candidates.\\
--C$_5$H$_{11}$NO (\textit{m/z} = 101): N,N-Diethylformamide (LCI = 2; FSRM = 15~\%) has suitable M and M-CH$_3$ signals. Its base peak is the resonance-stabilized CH$_4$N fragment ion on \textit{m/z} = 30, which is accommodated in Fig. \ref{fig:CHN}. N,2-Dimethylpropanamide, which could explain M and \mbox{M-CH$_3$}, too, yields too much C$_2$H$_4$NO on \textit{m/z} = 58. N,N-Dimethylpropanamide is an appropriate alternative candiate. Given the high FSRM of the formamide derivative, the presence of an additional parent species seems rather likely. Based on the non-observation of the M-H signal and the already fully explained CH$_4$N fragment ion, all other isomers with a relevant M signal have been ruled out.\\
--C$_6$H$_9$NO (\textit{m/z} = 111): Trimethyloxazole (LCI = 2; FSRM = 7.0~\%) is a suitable candidate because of its M, absent \mbox{M-H}, and M-CH$_3$ signal. The isoxazol isomer shows a strong \mbox{M-CH$_3$} signal, which makes it less favourable. A good alternative is 1-ethenyl-2-pyrrolidinone, which only yields one strong signal beside the M signal, namely, a CHO-fragment (C$_3$H$_4$O) on \textit{m/z} = 56 from the ring-opening reaction. Notably, there are many isomers that do not have reference data.\\
--C$_6$H$_{11}$NO (\textit{m/z} = 113): 1-Acetylpyrrolidine (LCI = 3; FSRM = 8.4~\%) yields M and M-Ac but no M-H and no M-CH$_3$. The M-Ac fragment ion shows in Fig. \ref{fig:CHN} on \textit{m/z} = 70. Unlike with the iminium fragment, the acetyl fragment ion can be accommodated within the signal's error margins as the signal has a very high intensity (see Fig.~2 in \citet{haenni2023}). All other isomers with reference mass spectra and relevant M signals can be ruled out based on the M-H, the M-CH$_3$, and/or the iminium ion fragment that cannot be accommodated.\\
--C$_7$H$_{11}$NO (\textit{m/z} = 125): 2-Acetyl-1,4,5,6-tetrahydropyridine (LCI = 1; FSRM = 7.0~\%) yields M, negligible M-H, and no relevant M-CH$_3$. M-Ac on \textit{m/z} = 82 shows up in the Fig. \ref{fig:CHN}. On \textit{m/z} = 97 (C$_5$H$_8$NO), no signal has been measured, which disfavours 2-acetyl-3,4,5,6-tetrahydropyridine. 1-Acetyl-1,2,3,4-tetrahydropyridine, on the other hand, has a similar spectrum to 2-acetyl-1,4,5,6-tetrahydropyridine. Hence it could contribute to the observed intensity as its additional signal on \textit{m/z} = 83, likely the C$_4$H$_5$NO fragment after ring opening and subsequent loss of C$_3$H$_6$, can be accommodated within the observed signal's error margins. All other isomers with reference mass spectra and relevant M signals can be ruled out based on the M-H, the M-CH$_3$, the M-C$_2$H$_6$, and/or the iminium ion fragment that cannot be accommodated.

\subsection{Unexplained intensity}\label{subsec:unexplained}
The measured intensities that remain unexplained with our Occam's razor-based spectral deconvolution procedure, i.e. blanks in Figs. \ref{fig:CHN} and \ref{fig:CHNO}, are mostly attributed to species with a small number of hydrogen atoms. For the C$_n$H$_m$N species, these are: C$_3$N (\textit{m/z} = 50) or C$_4$H$_2$N (\textit{m/z} = 64). The former signal might be associated with the cyanoethinyl radical, first detected tentatively towards CW Leonis (Infrared Catalogue +10216; \citet{guelin1977}) and later confirmed towards TMC-1 \citep{mcguire2022}. There is no signal registered in the analysed DFMS dataset at the position of the next species in this carbon-chain series, the cyanobutadiynyl radical (C$_5$N), which might be explained by a decreasing abundance with an increasing carbon-number. In the ISM, this molecule has been detected towards TMC-1 by \citep{guelin1998}. For a thorough review of the carbon-chain chemistry in the ISM (see e.g. \citealt{taniguchi2024}). The underexplained C$_4$H$_2$N signal might be associated with M-H of cyanopropyne or cyanoallene or another structural isomer of the sum formula C$_4$H$_3$N. According to \citet{mcguire2022} and literature therein, cyanopropyne (also known as methylcyanoacetylene) and cyanoallene have both been detected towards TMC-1. Recently, also the 1-cyano \citep{cabezas2025} and 3-cyano propargyl radicals \citep{cabezas2021} were reported towards TMC-1.\\
Of the C$_m$H$_n$NO species, C$_2$HNO (\textit{m/z} = 55) is clearly underpopulated. This likely indicates an additional parent that yields an M but not an M-H signal. Possible structures with some chemical relevance are cyanoformaldehyde, iminoethenone, nitrosoacetylene, ketoaziridine, or oxoacetonitrile. However, no EI-MS reference data for those species are available. Cyanoformaldehyde (IUPAC name: formyl cyanide) has been detected towards the star-forming region Sagittarius B2(N) with the Green Bank Telescope \citep{remijan2008}. A very minor but unaccounted for signal of C$_3$H$_8$NO on \textit{m/z} = 74 might be an M-H fragment of a C$_3$H$_9$NO parent that does not yield a relevant M signal. However, this signal's intensity is close to the instrument's detection limit and no reliable statement can be made.\\
These observations for unaccounted C$_m$H$_n$N and C$_m$H$_n$NO species are consistent with previous reports for 67P's pure hydrocarbons \citep{haenni2022} and O-bearing hydrocarbons \citep{haenni2023}, for which it has been suggested already that residual intensity is likely associated with radicals and other less stable and more reactive species without reference spectra. However, future systematic investigations of cometary data are needed to obtain a more comprehensive picture of hydrogen-deficient species.\\
In addition, some characteristic EI-fragments seem to be slightly underpopulated. One example is the typical, resonance-stabilized nitrile fragment NH--C--CH$_2$ on \textit{m/z} = 41, which results from McLafferty rearrangements in nitriles whose $\alpha$-carbon atom is not branched. However, also the two neighbouring signals show the same slight underpopulation which might rather indicate missing parent species. Considering the fact that the DFMS uses a 45 eV ionization voltage as compared to 70 or 75 eV used for the standard reference mass spectra collected in the NIST database, we expect a tendency that our instrument yields relatively less fragments. Consequently, stacking of NIST spectra might result in a systematic overpopulation of typical fragment signals. Indeed, we observe an overpopulation for certain characteristic fragments, for example the iminium ion on \textit{m/z} = 30 or the C$_2$H$_6$N fragment on \textit{m/z} = 44. A thorough calibration campaign may shine more light on the extent of this effect. Previous calibration work for a series of pure and O-bearing hydrocarbons, however, has shown that deviations of the DFMS spectra from the NIST ones are usually below or around 10\% of the relative signal intensity \citep{schuhmann2019a,schuhmann2019b}.

\section{Discussion}\label{sec:disc}
Here, we comment on our results and discuss methodological caveats and limitations of EI-MS. Furthermore, we review the different chemical functional groups observed and contextualize our findings by comparing them to other reservoirs of extraterrestrial N- and NO-bearing organics.

\subsection{General interpretation and caveats}\label{subsec:remarks}
Generally, the local abundance of molecules is expected to decrease with increasing molecular weight for combinatorial reasons (under assumption of a bottom-up chemistry) and for reasons of decreasing volatility. This means that more complex species are not only expected to form less abundantly due to longer formation pathways, but also to be less detectable via gas phase as they do not easily sublimate anymore. This is well consistent with our observations of decreasing signal intensities and increasing number of gaps in homologous peak series in the investigated dataset overviewed in Fig.~1 in \citet{haenni2022}. However, the C$_m$H$_n$N and C$_m$H$_n$NO species appear to be rather exceptional. While the C$_m$H$_n$N species have an intensity dip when $m = 4$, the C$_m$H$_n$NO species rather have an intensity peak when $m = 4$. In the following, we attempt a discussion of this observation:\\
For the C$_m$H$_n$N signals, our Occam's razor-based solution, presented in Sect. \ref{sec:res}, suggests that signals of species with $m \leq 4$ are mostly associated with chain-based molecules with preferably amine or nitrile functionality (Fig. \ref{fig:CHN} shades of red) while species with $m > 4$ are favourably associated with heterocycles (Fig. \ref{fig:CHN} shades of green and blue). Obviously, a minimum number of three atoms is needed to form a ring. The stability of cyclic species is influenced by factors including ring strain and aromaticity. Ring strain is commonly understood as a combination of angle strain and torsional strain. The angle strain is minimal when the molecular orbitals of a molecule overlap in a favourable way, which is for instance the case in 6-membered rings. Aromaticity generally increases the stability of a molecule as it lowers the energy of the molecular orbitals. Notably, the N atom can contribute to aromaticity with its electron lone pair. This property seems to be more pronounced in 5-membered rings as compared to 6-membered rings \citep{zong2020}. Our solution comprises several pyrrole and pyridine derivatives, which are aromatic 5- and 6-membered N-heterocycles with contribution of the N lone pair electrons. Generally, our solution suggests that the N atom is favourably located within the ring to form a heterocycle rather than an N-functionalized carbocycle which could be due to said stability gain. While our solution features 3-, \mbox{5-,} and 6-membered rings, it does not contain any 4-membered rings. It must be said, though, that for the 4-membered N-heterocycles with a double bond, 1- and 2-azetine, reference spectra are not available. Azetidine, without a double bond, is not a favourable candidate based on its fragmentation pattern. The methylated 3-membered heterocycles yield a better fit to the data. Why 3-membered but not 4-membered heterocycles should be observed is chemically not obvious. However, \citet{chuang2024} recently reported the tentative detection of a 3-membered ring, cyclopropenylidene, in several comets. This supports the evidence for 3-membered rings in this dataset, both in the form of 3-membered carbocycles \citep{haenni2022} and of 3-membered heterocycles (this work). In addition, we note an analogy between the methylated 3-membered and the methylated 5-membered heterocycle, which, according to our solution, both seem more likely than the respective 4-membered and 6-membered symmetrical heterocycles. This could imply preferred formation of the chiral isomer.\\
The picture we find for the C$_m$H$_n$NO species is similar when it comes to the position of the heteroatoms N and O. Both, the N and the O atom, seem to favourably participate in ring-formation, which yields molecules such as oxazole and isoxazole derivatives. Furthermore, our solution also suggests the presence of O-functionalized N-heterocycles, while the inverse is less compatible with the observations. The reason for the peak in signal intensity for C$_m$H$_n$NO species where $m = 4$, however, is unclear. The best candidates for this group of signals are methylated heterocycles or functionalized carbocycles. Alkylation on both chains and cycles seems to be abundantly present. Especially methylated chains and rings are frequently the most suitable candidates. We note that the solution with a methylated 5-membered heterocycle (where both N and O are in the ring) is more favourable than the solution with a 6-membered heterocycle (where both N and O are in the ring). Ethylation appears to be less common and propylation is not represented at all according to our solution. This result is intuitively compatible with a bottom-up formation of the observed ensemble of organics.\\
While our Occam's razor-based approach provides fundamental insights into cometary molecular complexity, the existence of isomers leads to a degeneracy in the parameter space, and the following methodological caveats require thorough consideration: First, the obtained solution is non-unique as there is some entanglement. The most striking example is the M/M-H ratio, which we often used to determine the most suitable candidate to explain the observed M and M-H parent signals. However, this ratio may not always be a good indicator as fragments of larger species, most frequently M-CH$_3$ fragments, could interfere. We indicated and discussed these cases in Sect. \ref{sec:res} above. On the other hand, there are also helpful cases where, for instance, the non-observation of M-H or M-CH$_3$ fragments can be used to rule out certain isomers. Especially for larger COMs, certain isomers can be clearly ruled out, while still more than one isomers remain compatible with the observations.\\
Second, some groups of molecules are not favourably identified via EI-MS of complex gas mixtures as they do not yield strong M signals or no M signals at all. Our analysis, however, requires the presence of an M signal for an unambiguous identification as we must explain the signals of the observed parent species. Moreover, fragments alone do not provide enough constraints as they can always be associated with various parents. Species with missing or weak M are for example amines, especially primary, linear ones with three or more carbon atoms, irrespective of where the amine function is located. But also carboxamides usually have fragment signals as base peak. For most radical species no reference mass spectra are available and we do not know whether or not these species yield strong M signals. Despite the fact that molecules showing weak or absent M signals (i.e. only fragment signals) might evade identification, their total contribution cannot exceed the fraction of the fragment signals that remains unexplained in our solution, plus the error on the observed signals. Most chemical functional groups yield characteristic fragments, such as \textit{m/z} = 30, 44, and 58 for chain-based amines, which should be underpopulated if parent molecules (and their fragments) are missing. Above, in Subsect. \ref{subsec:unexplained}, we have discussed that such characteristic fragments rather tend to be overpopulated and not underpopulated (with the exception of the nitrile fragment on \textit{m/z} = 41). On the other hand, cyclic and especially aromatic species tend to produce strong M and M-H signals and hence are easily identifiable with EI-MS.\\
Third, Occam's razor might not capture the true diversity of cometary organic matter, especially at higher masses as it aims to minimize the number of candidates used for the solution. For example, one ring molecule (explaining M and M-H simultaneously) is favoured over a combination of two chain-based molecules (explaining M and M-H separately). This might explain the observed underabundance of nitriles (which often do not yield M but just M-H) as inferred from the slight underpopulation of the typical nitrile fragment (NH-C-CH$_2$) at \textit{m/z} = 41. However, the chosen approach is the only way to identify specific molecules in the cometary coma through spectral deconvolution. Alternatively, statistical approaches, as pursued for instance in \citet{ruf2022} or \citet{varmuza2020}, only describe the properties of the whole ensemble of organics. While the application of such methods allows the extraction of unbiased information on the organic complexity, we are convinced that our Occam's razor-based method has its own merit as it allows us to place constraints on many individual candidate species. Moreover, we complemented Occam's razor with naive bottom-up chemistry considerations. In a few cases, we thus favoured a solution with two candidates over a solution with just one candidate if that one candidate would have an unreasonably high FSRM. Where we found that multiple isomers cannot be distinguished based on their fragmentation patterns and the selected candidate shows an unreasonably high FSRM, we stated that its FSRM probably should be understood rather as a total FSRM of multiple isomers.

\subsection{Contextualization}\label{subsec:context}
In the following, we strive to embed our findings into the bigger picture of the evolution of organic material in space -- from its cradle in the ISM towards the biochemistry of carbon-based life on Earth. The role of comets in this picture is two-fold:\\
(1) Comets contain some of the most pristine Solar System matter and it has been shown that they share many low-mass volatile species with astrophysical environments in the ISM, including N-bearing ones such as hydrogen cyanide, acetonitrile, cyanoacetylene, 2-propenenitrile, isocyanic acid, or methyl isocyanide \citep{bockelee-morvan2000,drozdovskaya2019,lopez-gallifa2024}, even in similar abundances relative to methanol. As of mid-2021, \citet{mcguire2022} listed more than 50 N-bearing neutrals in the census of interstellar molecules. These are mostly radicals, nitriles, imines, and a few amides. A detailed understanding of the cometary ices is a crucial point of reference for ISM studies and vice versa. This also holds for more altered Solar System matter: The inter-comparison especially to C-type asteroids, containing original organic matter that underwent significant post-processing including liquid water chemistry, allows us to constrain chemical evolution. Until today, very limited evidence exists for the occurrence of aqueous alteration in comets \citep{lisse2006,berger2011}.\\
(2) Alongside interplanetary dust, meteorites, and asteroids, comets may have delivered substantial amounts of prebiotic species to the early Earth through impacts and hence may have helped to kick-start life (see e.g. \citet{rubin2019a} and references therein). As N-bearing species are key players in prebiotic chemistry and eventually are incorporated in many biotically relevant molecules such as amino acids or nucleobases, they are a major target of the astro- and cosmochemistry communities. It has to be stated that molecular, i.e. soluble, organics from meteorites \citep{schmitt-kopplin2010} as well as from samples returned from asteroids \citep{schmitt-kopplin2023,glavin2025} expose an extremely high molecular diversity, which is missed in targeted searches. Untargeted characterizations of the organic chemical complexity were also done for analog materials prepared in the laboratory, for example in ice irradiation experiments \citep{ruf2019a,ruf2022,kipfer2024} or in nebulotron-synthesized N-containing samples \citep{leveque2024}.\\
In Table \ref{tab:summary}, we summarize our findings by indicating the total FSRM of the various functional groups as determined from our Occam's razor-based deconvolution approach. The total FSRM can be viewed as an estimate of the order of the magnitude of the abundances ratios. We group the molecules represented in our solution into different categories (cyclic versus acyclic; chemical functionality; organic versus inorganic). Before comparing our findings with both asteroids and meteorites as well as the ISM, we comment on the absolute relevance of the C$_n$H$_m$N and C$_n$H$_m$NO species. Therefore, we compare organic molecules with N to inorganic molecules with N observed with the DFMS. The most abundant N-bearing compound is ammonia (NH$_3$), followed by hydrogen cyanide (HCN) or its isomer (HNC), isocyanic acid (HNCO) or its isomer (HOCN), molecular nitrogen (N$_2$), and nitric oxide (NO). Note that because of its chemical properties, which are similar to those of inorganic cyanides, we follow the convention to assign HCN and HNC to the inorganic compounds, despite the fact they do contain a CH-bond, unlike HCNO or HOCN, among others. Nitrogen dioxide (NO$_2$), another inorganic candidate species, would be hidden under the CO$^{18}$O isotopologue peak of CO$_2$, but its abundance is probably negligible. For an investigation of the bulk abundances of these common cometary volatiles in 67P, we refer to \citet{rubin2019b}. In the dataset investigated for this work, NH$_3$ is significantly more prominent than methanol (FSRM = 172~\%). \citet{rubin2019b} reported NH$_3$ to be roughly three times more abundant relative to water than methanol around the comet's inbound equinox when the come conditions were relatively stable. Especially in Jupiter family comets, to which also comet 67P belongs, the NH$_3$ abundance relative to water seems to show a high variability \citep{dellorusso2016}. \citet{altwegg2020} have presented evidence for several ammonium salts sublimating from 67P's dust and have argued that semi-volatile ammonium salts may be a reason for the heliocentric distance dependence of the NH$_3$/H$_2$O ratio. In data presented here, the inorganic N-bearing compounds, including NH$_3$ as the most relevant volatile N carrier, make up for a total FSRM of 301~\% while organic N-bearing ones amount to a total FSRM of 292~\%. This suggests that organic C$_n$H$_m$N and C$_n$H$_m$NO species roughly equal the inorganic ones, at least when the cometary outgassing is including enhanced sublimation from ejected dust particles as is the case on 3 August 2015. Notably, a few organic species with multiple N or O atoms shown in Fig.~1 in \citet{haenni2022} are not included in this estimation as they will be topic of future investigations. In terms of total FSRM, these species do not substantially change the balance between organic and inorganic N carrieres as it presents itself here, though.\\
Our findings impact the apparent cometary N deficiency, which, for comet 1P/Halley, was reported to be about a factor 5 compared to the Solar System N abundance value (see Fig.~1 in \citet{geiss1987}). Geiss used an N abundance values for the Solar System from \citet{anders1982}, but recent updates are available \citep{lodders2025}. The total N abundance estimation for Halley included an amount of N from cometary volatiles observed at that time by the ion mass spectrometer aboard ESA's Giotto spacecraft \citep{balsiger1986}, mainly associated with HCN and NH$_3$, and an amount of N from cometary dust measured by a dust impact-ionization mass spectrometer on the Soviet Vega-1 space probe \citep{jessberger1988}. If now the gas abundance of N roughly doubles due to the here reported reservoir of organic N-bearing volatiles, this raises the total N abundance by a factor of about 1.5 and brings it closer to the Solar System value. If an additional reservoir of semi-volatile ammonium salts is considered, with evidence for 67P reported by \citet{altwegg2020} and \citet{poch2020}, the total cometary N abundance might be similar to the Solar System one. However, this estimation is based on FSRM and hence affected by differences in the ionization cross-sections. Most likely, it overestimates the heavy COMs as compared to the lighter molecules as the former tend to have larger cross-sections (see e.g. \citet{beran1969} and \citet{kumar2019}). Consequently, the importance of heavy COMs as a cometary N reservoir might be less pronounced as suggested in the above estimation. The N abundance, as well as its distribution between volatiles and refractories phases, including salts, is also topic of investigation for asteroids \citep{schmitt-kopplin2023,pilorget2024,glavin2025} and may be a key to understanding post-accretional processing (see  below).

\begin{table}[h]
   \caption[]{Total fragment sum relative to methanol (FSRM) of the different groups of volatiles from this work.}
   \centering 
   \begin{tabular}{ll}
   \toprule
   Functionality        & total FSRM (\%)             \\
   \midrule
   C$_m$H$_n$N          & 235                         \\
   C$_m$H$_n$NO         & 152                         \\
   \hline
   Acyclic              & 245                         \\
   Cyclic               & 142                         \\
   -- N-Heteroycles     & 89 (O-functionalized: 36)   \\
   -- NO-Heteroycles    & 41                          \\
   -- Carbocycles       & 12                          \\
   \hline
   Amines               & 39 (Methylamine: 17)        \\
   Nitriles             & 129 (Hydrogen cyanide: 81)  \\
   Amides               & 53 (Formamide: 12)          \\
   Lactams              & 20                          \\
   Oxazoles \& isoxazoles & 41                        \\  
   Others$^{(a)}$       & 105                         \\ 
   \hline
   Inorganic$^{(b)}$    & 301                         \\
   Organic$^{(c)}$      & 292                         \\
  \bottomrule
     \end{tabular}
  \label{tab:summary}
  \tablefoot{$^{(a)}$ Others include the following less abundant functionalities: imines, isocyanides, isocyanates, and acetylated N-heterocycles. We also include all unfunctionalized N-heterocycles in this residual group, even though some would actually classify them as amines. $^{(b)}$ To obtain the total amount of inorganic N, we sum over the FSRM of HCN and HNCO (both from this evaluation) and add NH$_3$ (FSRM = 172~\%), N$_2$ (FSRM = 17~\%), and NO (FSRM = 17~\%). $^{(c)}$ To obtain the total FSRM of organic N, we deduce from the total C$_n$H$_m$N and C$_n$H$_m$NO FSRM (387~\%) the fraction assigned to HCN (FSRM = 81~\%) and HNCO (FSRM = 14~\%).}
\end{table}

\subsubsection{Comparison to meteorites and asteroids}
Comparisons of our findings from cometary in situ data to those from analysis of collected meteorite samples or returned asteroid samples in laboratories on the ground must be made carefully. Chondrite analysis, for instance, involves solvent extraction and especially the use of acids leads to the hydrolysis of certain functional groups such as amides or lactams \citep{burton2012}. Consequently, the studies without acid extraction step in the sample work-up protocol should yield results that are more comparable to ours. Nevertheless, chemical reactions in the liquid phase during extraction, even if the solvent is simply water, are difficult to rule out or constrain. For the Murchison meteorite, a well-studied CM-type chondrite, \citet{jungclaus1976a} report 8~$\mu$g aliphatic amines per gram meteorite while \citet{aponte2014} reported 106 nmol/g free and 157~nmol/g acid hydrolyzed aliphatic amine content. Acid extraction increases the amine content as it hydrolyzes functional groups (e.g. amides) into amines \citep[e.g,][]{sephton2002}. The reported numbers indicate an increase by a factor of about 1.5. The biggest contribution is methylamine, which is roughly five times more abundant than ethylamine, the second most abundant amine. In comet 67P, we identified the following aliphatic amines with reasonable to high certainty: methylamine (FSRM = 17~\%), ethylamine (FSRM = 13~\%), 2-propanamine (isopropylamine; FSRM 1.1~\%), N,N-dimethyl-methylamine (FSRM = 3.4~\%), and N-methyl-2-propanamine (FSRM = 3.8~\%). The FSRM decrease from methylamine to ethylamine seems less pronounced. While it is likely that the primary amine isomer is more abundant than the secondary one, as has been discussed in Sect. \ref{sec:res}, their true ratio is difficult to constrain with mass spectrometry alone. \citet{glavin2008} reported the presence of methylamine and ethylamine for acid-hydrolyzed hot water extracts of the cometary matter returned from comet Wild 2 by the Stardust mission \citep{brownlee2006}. However, there was no significant detection of the free form, indicating, that these compounds were rather bound in some hdyrolyzable precursor. In the Murchison meteorite, the primary amine ethylamine was reported to be almost a factor of 5 more abundant than its secondary amine counterpart dimethylamine \citep{aponte2014}. Generally, secondary amines seem to be significantly less abundant than primary ones in this meteorite \citep{aponte2014}. For comet 67P, our solution suggests a similar conclusion: As we have argued above, the iminium ion on \textit{m/z} = 30 (proxy for primary amines), the C$_2$H$_6$N fragment on \textit{m/z} = 44 (proxy for secondary amines), and the C$_3$H$_8$N fragment on \textit{m/z} = 58 (proxy for tertiary amines)  allow us to place constraints on the total abundance of the primary, secondary, and tertiary amines, respectively. We observe, that the iminium ion is more abundant than the C$_2$H$_6$N fragment by approximately a factor of 3 but more abundant than the C$_3$H$_8$N fragment by only a factor of about 2. Notably, a direct comparison of the intensities of these characteristic amine fragments is not subject to biases of the chosen deconvolution strategy and applicable to any mixture of COMs analysed by EI-MS. A total of 25 aliphatic amines up to C$_6$ have been found in the Murchison meteorite \citep{aponte2014} while in the Orgueil meteorite, a CI-type and hence more aqueously altered chondrite, only 12 aliphatic amines up to C$_4$ could be detected \citep{aponte2015}. Antarctic carbonaceous chondrites seem to possess the same suite of aliphatic amines as the Murchison meteorite \citep{aponte2016}.\\
In fact, also N-heterocycles classify as secondary or tertiary (if the N is bonding to C only) amines, respectively. While we are planning to investigate and discuss nucleobases in a future study, we focus on simple N-bearing heterocycles with one N here: Over many decades, basic N-heterocycles have been searched for and studied extensively in meteorites, especially since the fall of the Murchison meteorite in 1969 \citep{folsome1971}. Murchison meteoritic matter contains a series of purines, pyrimidines, quinolines, and pyridines, including alkylated derivatives \citep{sephton2002}. A recent review of the extensive meteoritic inventory of N-heterocycles by \citet{martins2018} shows a diversity ranging from pyrrolidine derivatives (5-membered rings), to purine (annulated rings), and lactam representatives (up to 8-membered rings). In our deconvolution solution, N-heterocycles make up for a total FSRM of 89~\% (36~\% are additionally O-functionalized), while NO-heterocycles make up for a total FSRM of 41~\%. Including another 12~\% attributed to carbocycles, a bit over a third (i.e. 142 out of 387~\%) of the total observed C$_n$H$_m$N and C$_n$H$_m$NO signal intensity is attributed to cyclic species, the rest to acyclic ones. \citet{aponte2014} also reported the presence of the cyclic secondary amine pyrrolidine for the first time in meteoritic matter (from Murchison). While the evidence for the presence of pyrrolidine is insufficient, our data contains evidence for its dehydrogenated forms, i.e. pyrrolines and pyrroles, and potentially for its acetylated form. Our work indicates the presence of 3-, 5-, and 6-membered N-heterocycles, often alkylated or dehydrogenated or even aromatic. For 4-membered rings we find less compelling evidence, but they may be present at the expense of some methylated aziridine derivatives. These findings from this work are consistent with reports for 67P's pure carbocycles, where the 4-membered carbocyles seem to be less abundant than the 3- and 5-membered ones \citep{haenni2022}. The reason, however, is unclear as the ring strain is largest for the 3-membered rings. Possible synthetic pathways towards N-heterocycles in meteorites originate in an ammonium cyanide chemistry \citep{botta2002,callahan2011}, but also hexamethylenetetramine (HMT) or urea are high-ranking precursors (see e.g. \citet{oba2023a} and references cited therein). Notably, ammonium cyanide (NH$_4$CN) has been reported to sublimate from 67P's dust \citep{altwegg2020}. Extensive review work on the study of carbonaceous chondrites is available \citep{cronin1993,botta2002,sephton2002,glavin2018}.\\
An additional unknown in meteorite studies is terrestrial weathering that may occur prior to collection and curation to variable extents. Therefore, samples collected from near-Earth asteroids, such as asteroid (101955) Bennu and asteroid (162173) Ryugu, returned to Earth and kept in curation facilities under nitrogen fluxes provide valuable complementary insights into organics. According to its mineralogy, Ryugu seems to be associated with CI chondrites (with much less water content, though), while for Bennu the case is less clear cut. From remote sensing, Bennu appeared to be a CM-like asteroid while from laboratory analyses it rather seemed similar to Ryugu and CI chondrites. Ryugu samples analysed for small soluble organic molecules showed the aliphatic amines methylamine, ethylamine, \textit{n}-propylamine, and isopropylamine \citep{naraoka2023}. No dimethylamine was reported. The individual abundance was decreasing with increasing chain length while the total aliphatic amine abundance was an order of magnitude less than in the Orgueil chondrite \citep{parker2023}. New results from similar analyses of samples returned from asteroid Bennu showed that the C1-C6 amine distribution follows the trend of decreasing concentrations with increasing carbon number as well \citep{glavin2025}. However, the reported abundances were higher overall and a larger diversity of amines was observed as compared to Ryugu which was suggested to be a result of less extensive aqueous alteration processes on Bennu's parent body. Unlike Bennu, where isoproylamine and n-propylamine are present in similar abundances, the branched isomer of C$_3$H$_9$N seems to be more abundant than the straight-chain isomer in Ryugu by almost an order of magnitude. \citet{parker2023} suggested the dominance of the branched isomer of C$_3$H$_9$N in Ryugu to be the result of either favoured formation in a radical reaction or enhanced thermal stability, as branched molecules are less prone to thermal degradation. We find that the straight-chain isomer is less compatible with the analysed DFMS data as its strong iminium ion fragment cannot be accommodated in our solution, or only a minor amount of it, if some of the ethylamine is replaced with dimethylamine (see Sect. \ref{sec:res}). We thus cannot place stringent constraints on the branched-versus-straight ratio in C$_3$H$_9$N. \citet{aponte2014} reported the branched C$_3$H$_9$N isomer to be more abundant in the Murchison chondrite than the straight-chain one by about a factor 1.5 in the free form. The difference was found to be less pronounced in the acid-hydrolyzed extract.\\
Similar to meteorites, also the samples returned from asteroids seem to possess a diversity of N-heterocycles: For Ryugu, alkylated derivatives of pyridine, piperidine, pyrimidine, imidazole, and pyrrole were identified in a methanol extract \citep{naraoka2023}. It was suggested that differences in the number of carbon atoms in the alkylpyridine homologs identified in Ryugu and CM chondrites, including Murchison, may be related to hydrothermal history of the parent bodies or irradiation conditions. In Bennu matter, an extensive set of N-heterocycles, including all five nucleobases of terrestrial life, have been identified \citep{glavin2025}. The total abundance of N-heterocycles in Bennu extracts amounts to approximately 5 nmol/g. This is up to an order of magnitude more than what was reported for Ryugu and was suggested to be the result of less pronounced aqueous alteration on Bennu's parent body. The degree and effects of aqueous alteration and the related chemical implications for the organics budget of both these C-type asteroids, Bennu and Ryugu, have not been fully understood until today. However, finding the same main N-bearing species among comet 67P's volatile organics and in Ryugu, Bennu, and carbonaceous chondrites might be an indication that actually many of those species could have formed prior to planetesimal formation. Substantial work will be needed, though, to systematically investigate the abundances of individual cometary COMs and especially of different isomers, which could be realized for instance in the framework of a cometary sample return mission featuring a gas-chromatography-coupled mass spectrometer. This would pave the way towards an enhanced understanding of the post-accretional processing of organics and its effects on the chemical diversity and complexity.

\subsubsection{Comparison to the ISM}
While organic amines and N-heterocycles are good candidates to explain the mass spectra analysed for this work and are abundantly present in meteorite and asteroid matter, this does not seem to be the case for the ISM. Until today, only methylamine has been repeatedly identified in the ISM \citep{mcguire2022} and more recently a detection of vinylamine was reported towards the Galactic center cloud G+0.693-0.027 \citep{zeng2021}. The latter authors also reported a tentative identification of ethylamine. The detection of N-heterocycles has remained elusive until today. \citet{mcguire2022} mentions the tentative detection of aziridine. Aziridine, pyrrolidine, and piperidine (including dehydrogenated and alkylated forms of those molecules) provide an Occam's razor-compatible explanation for the DFMS data analysed in this work.\\
Notably, molecules must possess a dipole moment in order to be detectable via rotational spectroscopy, the technique commonly used to probe the gas phase in the ISM. Consequently, some groups of molecules are more easily identified than others and some, for example unfunctionalized symmetric hydrocarbons, are hardly detectable, for example unfunctionalized polyynes (carbon chains) or many aromatic ring systems. A work-around is the probing of such symmetric hydrocarbons via CN-functionalized analogs. It has been proven practical to use the molecules  benzonitrile \citep{mcguire2018a} and cyano-naphthalene \citep{mcguire2021}, among others, as proxies for their unfunctionalized counterparts. The presence of a dipole moment is not required for the analytical method of EI-MS applied in this work. As discussed above, also EI-MS has its limitations, and hence, methodological differences are adding to the effective differences between cometary organics and organics in the ISM. For these reasons, common ISM targets are the CN-functionalized carbocycles. But from our study, CN-functionalized carbocycles are clearly less common in comets than N-heterocycles.\\
Low-mass nitriles and N-bearing acids are easily detectable by both EI-MS and rotational spectroscopy. A subset of such molecules, including multiple previously reported cometary species such as hydrogen cyanide, acetonitrile, cyanoacetylene, 2-propenenitrile, isocyanic acid, or methylisocyanate, has been previously shown to be present in both comet 67P and the ISM \citep{drozdovskaya2019,lopez-gallifa2024}. For imines, unfortunately, no reference mass spectra exist and we only infer the presence of methanimine under the assumption of similarity to the fragmentation behavior of formaldehyde. If ethanimine yields substantial M and M-H as well, then this molecule might be a viable (partial) replacement of aziridine. Some ISM environments, however, such as the molecular cloud G+0.693-0.027, appear to be particularly rich in imines. In the same cloud, methanimine, ethanimine, cyanomethanimine, and propargylimine have been detected \citep{bizzocchi2020}. The presence of imines in comet 67P's coma is therefore possible if not likely, but probably also with decreasing abundances as observed for the amines. The acquisition of reference EI mass spectra has the potential to place further constrains on the presence of cometary imines. Also, the hydrogen-deficient molecules, including radicals abundant in the ISM, have to be further investigated and constrained. As we have discussed above in Subsect. \ref{subsec:unexplained}, there is yet unattributed intensity in the DFMS sum spectrum analysed for this work that is compatible with common ISM species such as C$_3$N (potentially the cyanoethinyl radical), C$_4$H$_2$N (potentially cyanopropyne or cyanoallene or an isomer thereof), or C$_2$HNO (potentially cyanoformaldehyde or an isomer thereof). Future systematic work will allow us to place more stringent constraints on such species.

\section{Summary and conclusions}\label{sec:sumconc}
In this work, we  presented and discussed a non-unique Occam's razor-based spectral deconvolution solution of an organic-rich EI-MS sum spectrum acquired by the DFMS in comet 67P's coma on 3 August 2015, focusing on the C$_n$H$_m$N and C$_n$H$_m$NO species. Occam's razor was combined with abundance constraints compatible with a naive bottom-up chemistry as well as information from studies of other relevant astrophysical environments. This approach led us to the following conclusions:

\begin{itemize}
	\item As already reported for comet 67P's pure hydrocarbons \citep{haenni2022} and for its O-COMs \citep{haenni2023}, the observed signals of C$_n$H$_m$N and C$_n$H$_m$NO species are best reproduced by a chemically diverse ensemble of cometary COMs similar to meteoritic and asteroidal soluble organic matter. We have found evidence for various chemical functionalities including amines, nitriles, and heterocycles, the ratio of acyclic to cyclic species being roughly 2 to 1. We  summed up the FSRM values in Table \ref{tab:summary}, which suggests that  inorganic and organic nitrogen carriers may be similarly abundant in total.
	\item Cyclic species, which often produce strong M and M-H signals, are more easily identified than chain-based species, which often produce either strong M or M-H. Occam's razor suggests that the solution with the smaller number of molecules is the preferable one, which appears debatable in face of the growth of combinatorial possibilities of structural isomers with a growing number of atoms in the molecule. However, characteristic fragment signals, such as the typical amine fragment on \textit{m/z} = 30, can be used to constrain the contribution of such hard-to-detect molecules. These characteristic fragment signals tend to be overpopulated in our solution. This means that while additional contributions of hard-to-detect species are highly likely, their abundance must be limited by the unaccounted for intensity plus the error margins on the observations.
	\item The typical primary amine fragment  (CH$_4$N on \textit{m/z} = 30) is approximately three times more abundant than the typical secondary amine fragment (C$_2$H$_6$N on \textit{m/z} = 44), but is only two times more abundant than the typical tertiary amine fragment (C$_3$H$_8$N on \textit{m/z} = 58). These fragments reliably constrain the total abundance of primary, secondary, and tertiary amines in an ensemble of COMs without going though full deconvolution of the dataset.
	\item N-heterocycles seem to be more compatible with the observations than N-functionalized carbocycles, and they are good candidates for the signals observed, especially in the higher-mass part of the DFMS spectrum. In NO-bearing species, both NO-heterocycles (e.g. oxazoles and isoxazoles) and O-functionalized (e.g. acylated) N-heterocycles are observed. From our study, there is no evidence for N-functionalized O-heterocycles.
	\item Alkylation seems to be very common in both chain-based and cyclic species in the form of short side-chains. For example, and in agreement with reports from meteorite and asteroid Ryugu studies, isopropylamine more closely matches the observed signals than its unbranched counterpart, which would lead to further overpopulation of the iminium ion fragment on \textit{m/z} = 30.
	\item The presence of amino acids and nucleobases falling into the class of C$_n$H$_m$N$_x$O$_y$ species, where $x + y > 2$, will be investigated separately in the future. It will be challenging to clearly identify such molecules in the complex mixture of organics present in 67P's coma. But it may be possible to derive meaningful upper limits from the absence of signals that should be clearly detectable according to EI-MS reference data.
\end{itemize}

The set of molecules used in our approach, together with their isomers where applicable, may inform future searches in related astrophysical environments. We achieved the clear identification of many smaller N- and NO-COMs, but the true chemical diversity for larger COMs, where the parameter space becomes increasingly degenerate, is probably underestimated. While we consider residual minimization as being too limited and the unambiguous identification of large COMs from DFMS sum fragmentation patterns to be impossible, we see more potential for instance in multivariate statistical methods or chemical networks to characterize the ensemble of cometary organic matter in unbiased ways. Particularly for larger COMs, top-down synthetic pathways may also play a role. This renders collaborative efforts among modelling, laboratory, and observational communities essential to deepen our understanding of the origin and evolution of organic complexity, from the ISM towards emerging and evolved planetary systems.

\begin{acknowledgements}
We gratefully acknowledge the work of the many engineers, technicians and scientists involved in the Rosetta mission and in the planning, construction, and operation of the ROSINA instrument in particular. Without their contributions, ROSINA would not have produced such outstanding data and scientific results. Rosetta is an ESA mission with contributions from its member states and NASA. Work at the University of Bern was funded by the Canton of Bern and the Swiss National Science Foundation (200020\_207312). S.F.W. acknowledges financial support of the SNSF Eccellenza Professorial Fellowship PCEFP2\_181150. M.R.C. acknowledges support from NASA grants 80NSSC18K1280 and 80NSSC20K0651. J.D.K. acknowledges support from the Belgian Science Policy Office. Especially, we thank Prof. Dr. B. McGuire for sharing with us his ISM perspective on the work presented here.
\end{acknowledgements}

\bibliographystyle{aa}
\bibliography{literature}

\end{document}